%% file: main.tex
\documentclass[journal]{IEEEtran}

\ifCLASSINFOpdf

\else

\fi

\usepackage[utf8]{inputenc}
\usepackage{amssymb}
\usepackage{stfloats}
\usepackage{placeins}
\usepackage{physics}
\usepackage{overpic}
\usepackage{caption}
\usepackage{amsmath}
\usepackage{comment}
\usepackage{mathtools}
\usepackage{booktabs}
\usepackage{cite}
\usepackage{subfig}
\usepackage{lipsum}
\usepackage{graphicx}
\usepackage{textcomp,gensymb}
\usepackage{glossaries}
\usepackage{multicol}
\usepackage{algorithm}
\usepackage[noend]{algpseudocode}
\usepackage{tikz}
\usepackage[english]{babel}
\usepackage{amsthm}
\theoremstyle{plain}
\usepackage{dirtytalk}

\graphicspath{ {./images/} }

\usepackage{amsmath}

\usepackage{mathtools}

\usepackage{tabularx}
\newcolumntype{L}[1]{>{\raggedright\arraybackslash}p{#1}}
\newcolumntype{C}[1]{>{\centering\arraybackslash}p{#1}}
\newcolumntype{R}[1]{>{\raggedleft\arraybackslash}p{#1}}
\usepackage{multirow}
\usepackage{mathtools, cuted}

\addto\captionsenglish{}
\input{acronyms}

\newcommand{\AM}[1]{\textcolor{black}{#1}}

\begin{document}
\title{Reinforcement Learning-Based Policy Optimisation For Heterogeneous Radio Access}

\author{\IEEEauthorblockN{Anup~Mishra,  \IEEEmembership{Member, IEEE}, {\v C}edomir Stefanovi{\'c}, \IEEEmembership{Senior Member, IEEE}, Xiuqiang~Xu, Petar~Popovski, \IEEEmembership{Fellow, IEEE}, and Israel Leyva-Mayorga, \IEEEmembership{Member, IEEE} \vspace{-1cm}}

\thanks{The authors Anup Mishra, {\v C}edomir Stefanovi{\'c}, Petar Popovski, and Israel Leyva-Mayorga are with the Department of Electronic Systems, Aalborg University, Aalborg 9220, Denmark (e-mail:anmi@es.aau.dk; ilm@es.aau.dk; petarp@es.aau.dk). This work is funded by project WITS.}
\thanks{Xiuqiang Xu is with Huawei Technologies, Shanghai, China (e-mail:xuxiuqiang@huawei.com).}}

\maketitle
\begin{abstract}
Flexible and efficient wireless resource sharing across heterogeneous services is a key objective for future wireless networks. In this context, we investigate the performance of a system where latency-constrained \gls{iot} devices coexist with a broadband user. The base station adopts a grant-free access framework to manage resource allocation, either through orthogonal \gls{ran} slicing or by allowing shared access between services. For the \gls{iot} users, we propose a \gls{rl} approach based on double \gls{ql} to optimise their repetition-based transmission strategy, allowing them to adapt to varying levels of interference and meet a predefined latency target. We evaluate the system’s performance in terms of the cumulative distribution function of \gls{iot} users' latency, as well as the broadband user’s throughput and \gls{ee}. Our results show that the proposed \gls{rl}-based access policies significantly enhance the latency performance of \gls{iot} users in both \gls{ran} Slicing and \gls{ran} Sharing scenarios, while preserving desirable broadband throughput and \gls{ee}. Furthermore, the proposed policies enable \gls{ran} Sharing to be energy-efficient at low \gls{iot} traffic levels, and \gls{ran} Slicing to be favourable under high \gls{iot} traffic.
\end{abstract}
\glsresetall
\glsresetall
\vspace{-0.5cm}
\section{Introduction}\label{Intro}
\AM{Future wireless networks, including beyond \gls{5g} and \gls{6g} systems, are expected to support a broad spectrum of heterogeneous services such as \gls{iot} and broadband connectivity. These services come with stringent \gls{qos} requirements across diverse use cases, including smart cities, remote sensing, and \gls{v2x} communication, among others\cite{Elayoubi@Slicing}.  To efficiently accommodate these diverse demands, \gls{ran} Slicing has been widely regarded as a promising solution\cite{Elayoubi@Slicing}. By partitioning the network infrastructure into logical slices, \gls{ran} Slicing enables tailored access and resource policies for different service types\cite{Elayoubi@Slicing}. This allows each slice to operate with configurations best suited to its \gls{qos} requirements, thereby facilitating efficient coexistence of diverse applications\cite{Elayoubi@Slicing, Mohammed@RANSlicing_AIML}.}
\par \AM{To support dynamic and efficient management of slices, existing research has explored the use of \gls{rl} for optimising resource allocation in sliced \glspl{ran}\cite{Mohammed@RANSlicing_AIML}. These \gls{rl}-based methods are especially attractive due to their ability to adapt to stochastic environments and learn optimal strategies through interaction, without requiring exact system models\cite{Mohammed@RANSlicing_AIML,Ley23Asilomar,nasrin@ravi}. In particular, \gls{rl} has been extensively applied to inter-slice radio resource block allocation, intra-slice power control, scheduling, etc \cite{Mohammed@RANSlicing_AIML, Ley23Asilomar,nasrin@ravi}. Building on this line of work, recent effort has explored  optimisation of access policy using \gls{rl}, particularly for \gls{iot} users with grant-free access, within a sliced network context\cite{Ley23Asilomar}. The study in \cite{Ley23Asilomar} designed repetition-based transmission strategy of an \gls{iot} user to enable its coexistence with a broadband user.  However, the considered setup involved only a single \gls{iot} user, allowing for a model-based single-agent \gls{rl} formulation. Given the anticipated surge in \gls{iot} device density, addressing the coexistence challenge with multiple \gls{iot} users becomes imperative, especially as model-based solutions may no longer be tractable in such complex and dynamic environments~\cite{Mohammed@RANSlicing_AIML, Elayoubi@Slicing, nasrin@ravi,Ley23Asilomar}.}
\begin{figure}[!t]
    \centering
    \includegraphics[width=0.7\linewidth]{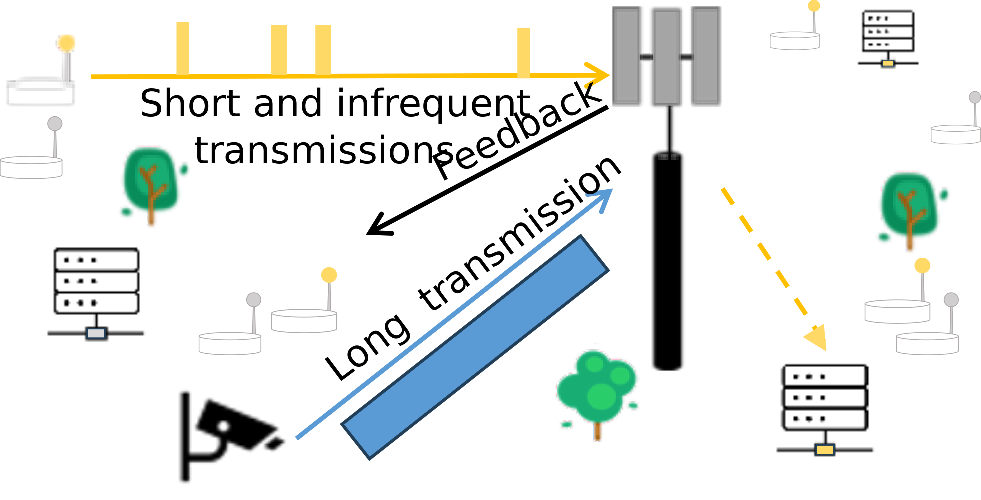}
    \caption{Uplink scenario with one user running a broadband service and multiple \gls{iot} users running intermittent services.}\vspace{-0.8cm}
    \label{fig:Uplink_Scenario}
\end{figure}
\par \AM{Motivated by the above discussion, this paper investigates the coexistence of a broadband user and multiple \gls{iot} users in a shared uplink scenario. The \gls{bs} slices its \gls{ran} resources between the two user types, where \gls{iot} users employ a repetition-based grant-free transmission mechanism. Upon activation, \gls{iot} users transmit packets without prior scheduling, and the \gls{bs} utilises the capture effect, along with both inter-slot and intra-slot \gls{sic}, to decode them~\cite{Giovvani@EIRSA}. This transmission and reception process is akin to that of \gls{irsa} with \gls{sic}\cite{Giovvani@EIRSA,Eleni@DD}. The objective is to optimise the \gls{iot} users' transmission policies under latency constraints and a fixed frame structure, with the aim of maximising overall system performance. Conventional \gls{irsa}-based studies typically optimise the repetition degree distribution to maximise throughput or minimise \gls{plr} under asymptotic assumptions\cite{Liva@DD,Eleni@DD}. However, such approaches are ill-suited for the considered setting, where finite frame lengths for latency guarantees are critical. To this end, we propose a \gls{rl}-based formulation. Given the stochastic and complex nature of the access environment due to multiple \gls{iot} users, model-based optimisation becomes intractable\cite{Mohammed@RANSlicing_AIML}. Therefore, we adopt a decentralised, model-free \gls{marl} framework, allowing each \gls{iot} user to independently learn its  strategy. System performance is evaluated in terms of the \gls{iot} latency \gls{cdf}, broadband throughput, and \gls{ee}, under both \gls{ran} Slicing and \gls{ran} Sharing regimes. Numerical results demonstrate that the proposed RL-based scheme significantly improves \gls{iot} latency while maintaining high throughput and \gls{ee} for the broadband user, thereby enabling their seamless coexistence.}
\vspace{-0.25cm}
\section{System Model}\label{Sysmod}
We consider an uplink scenario where a broadband user and \gls{iot} users are allocated sliced \gls{ran} resources to communicate with the \gls{bs}. Both the users and the \gls{bs} are equipped with a single antenna. The users are indexed as $m\in\mathcal{M}=\{b, i_{1},\,i_{2},\ldots,i_{J}\}$, where the broadband user is denoted by $m=b$ and the \gls{iot} users by $m=[i_{1},\,i_{2},\ldots,i_{J}]$, with the set of \gls{iot} users defined as $\mathcal{J}=\{1,\ldots,J\}$. The available wireless resources span a frequency band of bandwidth 
$B$ Hz \cite{Ley23Asilomar}, which is divided into three sub-bands: $1$) $B_{1}$, reserved for the broadband user, $2$) $B_{2}$, reserved for \gls{iot} users, and $3$) $B_{3}$, shared by both type of users, such that $B_{1}+B_{2}+B_{3}=B$. Let $\alpha_{m,w}\in\{0,1\}$ denote the allocation of user $m$ to sub-band $w\in\{1,2,3\}$, where $\alpha_{m,w}=1$ if user $m$ is assigned to sub-band $w$, and $\alpha_{m,w}=0$ otherwise. We consider a time-slotted communication system with slot duration $T_{s}$. The communication is segmented into frames, each consisting of $T_{F}$ consecutive time slots. All users are frame- and slot-synchronous. Subsequently, the two resource sharing scenarios considered in this paper are given by
\begin{enumerate}
    \item \textit{\Gls{ran} Slicing:} Users of different service types are allocated non-overlapping frequency sub-bands, with $B_{1}$ reserved for the broadband user and $B_{2}$ for \gls{iot} users. This allocation is defined by setting $\alpha_{b,1}=1$, $\alpha_{i_j,2}=1,\,\forall j\in\mathcal{J}$, $B_{3}=0$, and  $B=B_{1}+B_{2}$\cite{Ley23Asilomar}. 
    \item \textit{\gls{ran} Sharing:} All users, notwithstanding the service type, are allocated the entire bandwidth by setting $\alpha_{b,3}=1$, $\alpha_{i_j,3}=1,\,\forall j\in\mathcal{J}$, $B_{1}+B_{2}=0$ and $B_{3}=B$. 
\end{enumerate}
\vspace{-0.45cm}
\subsection{Transmission Model}\label{SysMod_TransMod}
\vspace{-0.05cm}
Within a frame, the first $T_{F}-1$ slots are allocated for uplink transmission, while the last slot is dedicated to downlink feedback, \gls{ack} or \gls{nack}, from the \gls{bs} to the users. The time slots and frames are indexed by  $t\in{\mathbb{N}}$ and $f\in{\mathbb{N}}$, respectively. Next, we outline the transmission policies for the broadband and \gls{iot} users.
\subsubsection{Broadband user}\label{SysModTransMod_BBU}
The user segments its data into packets and applies an ideal rate-less packet-level coding scheme to effectuate \gls{fec}, encoding blocks of $K$ source packets into linearly independent packets. An encoded block may span multiple frames, and the \gls{bs} decodes it upon successfully receiving $K$ encoded packets. At the end of each frame, the \gls{bs} provides feedback indicating successful or failed decoding. If an \gls{ack} is received, the user transmits the next source block; otherwise, it continues transmitting the current block. This transmission strategy guarantees the broadband user a reliability of $1$. \AM{We would like to highlight that the analysis can be simply extended to multiple broadband users, however, we focus on a single user for ease of exposition.}
\subsubsection{\gls{iot} users}\label{SysModTransMod_BBU}
The \gls{iot} users generate a new packet of length $L$ with probability $p_{j},\, j\in\mathcal{J}$ at each time slot. Without loss of generality, we assume $p_{j}=p_{a},\, \forall j\in\mathcal{J}$. The transmission queue of all \gls{iot} users is limited to a single packet, meaning any new arrivals are discarded if a previously generated packet is still in transmission. Packets are transmitted to the \gls{bs} in the next frame following a repetition-based grant-free access protocol using either sub-band $2$ or $3$. The number of repetitions of a packet transmitted in frame $f$, referred to as the repetition degree, is denoted by $a_{j},\,j\in\mathcal{J}$. Based on the feedback from the \gls{bs} at the end of a frame, \gls{iot} users adjust their transmission strategy for the next frame.
\vspace{-0.3cm}
\subsection{Physical Layer Model}
The channel coefficient between the \gls{bs} and user $m\in\mathcal{M}$ at time slot $t$, denoted as $h_{m,t} \in \mathbb{C}$, is modelled as a random variable incorporating both large-scale and small-scale fading. The small-scale fading follows a \gls{cscg} distribution with zero mean and unit variance, while the large-scale fading loss, denoted by $\beta_{m}=\mathbb{E}\{\lvert h_{m,t} \rvert^{2}\},\forall m\in\mathcal{M}$, is modelled as in \cite[eq. $(2)$]{Ley23Asilomar}. Next, the transmit signal between user $m$ and the \gls{bs} at time slot $t$ is denoted as $x_{m,t} \in \mathbb{C}$, while the transmission power of user $m$ is given by $P_{m,t} \in [0, P_{\max}],\,\forall m\in\mathcal{M}$. Consequently, the received signal at the \gls{bs} during uplink transmission in the $w$-th sub-band and $t$-th time slot can be expressed as:
\begin{equation}\label{eq:rx_signal}
    y_{w,t}= \sum_{m\in\mathcal{M}}h_{m,t}\,x_{m,t}\,\alpha_{m,w}+n_{w,t},
\end{equation}
where $n_{w,t}$ is the \gls{cscg} additive noise with zero mean and variance $\sigma_{w}^{2}$. Following \eqref{eq:rx_signal}, the \gls{sinr} for user $m$ in sub-band $w$ at time slot $t$ can be expressed as
\begin{equation}\label{eq:sinr_m}
    \gamma_{m,w,t}=\frac{\lvert h_{m,t}\rvert^{2}P_{m,t}\alpha_{m,w}}{\sum_{n\neq m,n\in\mathcal{M}}\lvert h_{n,t}\rvert^{2}P_{n,t}\alpha_{n,w}+\sigma_{w}^{2}},
\end{equation}
Subsequently, the probability of successfully decoding a packet transmitted by user $m$ in sub-band $w$ at time slot $t$ can be calculated as $p_{m,w,t}=1-\epsilon_{m,w}=\textrm{Pr}\left(\gamma_{m,w,t}< \gamma_{m,w}^{\textrm{min}}\right)$\cite{mishra2024coexist}. Here, $\gamma_{m,w}^{\textrm{min}}=2^{r_{m}/B_{w}}-1$ is the threshold for decoding the signal of user $m$ in sub-band $w$ at time slot $t$ as function of rate $r_{m}$, and $\epsilon_{m,w}$ is the error probability\cite{Ley23Asilomar,mishra2024coexist}. To decode the packets of different users, we consider a \gls{sic} decoder with \textit{capture}\cite{Giovvani@EIRSA}. The receiver aims to decode packets of as many users as possible per slot by leveraging the capture effect, which allows decoding of the strongest received signals. Once a packet is successfully decoded, its interference is first removed from the current slot (intra-slot \gls{sic}) and then from slots where its replicas were transmitted (inter-slot \gls{sic})\cite{Giovvani@EIRSA}.
\vspace{-0.1cm}
\section{Performance Optimisation}\label{Perf_Opt}
In this section, we optimise the access policy of the \gls{iot} users to maximise their latency-reliability performance using a decentralised \gls{marl} approach. Here, access policy optimisation is performed at the user side with no coordination among users. Decentralised learning is desirable as it reduces communication overhead and enhances scalability. Since \gls{rl} problem formulation requires defining a \gls{mdp} or \gls{pomdp}, we first model the access of an \gls{iot} user with a state space $\mathcal{S}_{j}$, an action space $\mathcal{A}_{j}$, and a reward function $\mathcal{R}_{j}$, $j\in\mathcal{J}$\cite{Ley23Asilomar,Mohammed@RANSlicing_AIML}. Further, at time slot $t$, the state of \gls{iot} user $j$ is defined by the tuple $s_{j,t}=\left(l_{j},v_{j},\delta_{j}\right)$, where $l_{j}$ denotes the latency of the current packet in the transmission queue, $v_{j}$ is the total number of repetitions transmitted since its generation, and
$\delta_{j}$ indicates whether the packet has been successfully decoded by the \gls{bs} ($\delta_{j}=1$) or not ($\delta_{j}=0$). At the start of each frame $f\in\{0,1,\ldots\}$, \gls{iot} user $j$ selects an action $a_{j}\in\mathcal{A}_{j}=\{0,\ldots,T_{F}-1\}$, determining the repetition degree for that frame. Repetitions are transmitted consecutively from the first time slot to minimise latency. If multiple repetitions are successfully decoded in the same frame, latency $l_{j}$ is set to the first successful reception; otherwise, it increases with each slot. The user partially observes the system state only at the end of the frame, after receiving feedback for its own transmission and before selecting the next action. A packet is removed from the queue upon successful decoding or when $l_{j}$ exceeds the maximum predefined latency, with the state resetting to $(0,0,0)$ in both cases. Finally, the reward for a give state $\left(l_{j},v_{j},\delta_{j}\right),\,\forall j\in \mathcal{J}$, is given by
\begin{equation}
    R_{j}\left(l,v,\delta\right)=\begin{cases} 
    \frac{50}{(l+1)^{2}+(v+1)}, & \text{if } \delta = 1 \\
    \max\left(-1,-0.03l-0.01v\right), & \text{if } \delta = 0.
\end{cases}
\end{equation}
Such a reward structure prioritises low latency while simultaneously minimising excessive transmissions. If the transition probabilities from a given state  $s_{j,t}\in\mathcal{S}_{j}$ at time $t$ to any given state  $s_{j,t+T_{f}}\in\mathcal{S}_{j},\,\forall T_{f}\in\{0,\ldots,T_{F}-1\}$ were known to user $j$, each \gls{iot} user could have employed \gls{vi} to determine their optimal transmission policy for the given reward structure\cite{Mohammed@RANSlicing_AIML}. In fact, \cite{Ley23Asilomar} applied this approach for a single \gls{iot} user, where the transition probability was fully characterised by $p_{m,w,t}$. However, with multiple \gls{iot} users, the system dynamics cannot be accurately modelled, making it intractable to analytically compute transition probabilities. Therefore, \gls{iot} users interact with the environment and optimise transmission using model-free \gls{rl}, specifically \gls{ql}. To this end, \gls{iot} users employ softmax exploration to balance the exploration-exploitation trade-off during training. The objective of each user is to update its respective $Q$-function with each interaction, using the tuple $(s_{j},a_{j},s_{j}^{'},R_{j})$, as the user transitions from state $s_{j}$ to $s_{j}^{'}$. The update rule for the $Q$-functions is given by
\begin{equation}
\begin{split}
    &Q_{j}^{1}(s_{j}, a_{j})\leftarrow (1-\mu)Q_{j}^{1}(s_{j}, a_{j}) + \mu( R_{j}(s_{j}) + \varphi Q_{j}^{1}(s_{j}^{'},a_{j}^{2}))\nonumber\\
    &Q_{j}^{2}(s_{j}, a_{j})\leftarrow (1-\mu)Q_{j}^{2}(s_{j}, a_{j}) + \mu( R_{j}(s_{j}) + \varphi Q_{j}^{2}(s_{j}^{'},a_{j}^{1}))\nonumber
\end{split}
\end{equation}
where $\mu$ is the learning rate, $\varphi$ is the discount factor, $a_{j}^{1}=\arg\max_{a_{j}^{'}} Q_{j}^{1}(s_{j}^{'},a_{j}^{'})$ and  $a_{j}^{2}=\arg\max_{a_{j}^{'}} Q_{j}^{2}(s_{j}^{'},a_{j}^{'})$\cite{Ley23Asilomar,Mohammed@RANSlicing_AIML}. Note that we employ \gls{doql}, a variant of \gls{ql} where each user maintains two $Q$-functions. Introduced to mitigate overestimation bias, \gls{doql} updates each $Q$-function using the next-state value from the other $Q$-function \cite{NIPS2010_091d584f}. This prevents the over-selection of actions with inflated values, a common issue in standard \gls{ql}, where the same $Q$-function is used for both action selection and evaluation, particularly in noisy or stochastic environments. \gls{doql} converges to the optimal policy in the limit, with details provided in \cite{conv@QL,NIPS2010_091d584f}. Subsequently, the optimal transmission policy for user $j\in\mathcal{J}$, $\pi_{j}^{*}(s_{j})$, is obtained by first averaging its two $Q$-functions as $Q_{j}^{*} = ({Q_{j}^{1} + Q_{j}^{2}})/{2}$ and then computing $\pi_{j}^{*}(s_{j}) = \arg\max_{a_{j} \in \mathcal{A}{j}} Q_{j}^{*}(s_{j}, a_{j})$.
\begin{table}[t]
	\caption{Simulation Parameters}\vspace{-0.2cm}
	    \label{tab:Sim_table}\centering
	\begin{tabular}{l l l}
		\toprule[0.4mm]
		\textbf{Parameter} & \textbf{Symbol}& \textbf{Value}\\
  		\toprule[0.4mm]
   Broadband user erasure probability & $\epsilon_{b}^{*}$& $0.1$\\
   Broadband user maximum data rate & $r_{b}^{\textrm{max}}$& $5$ Mbps\\
   Maximum transmission power & $P_{\textrm{max}}$ & $200$ mW \\
   Antenna gains & $G_{t}, G_{r}$& $10$\\
   Time slot duration & $T_{s}$ & $1$ ms\\
   Carrier frequency &$f_{c}$& $2$ GHz\\
   System bandwidth & $B$ & $1$ MHz\\
   Noise temperature & $T_{w}$& $190$ K\\
   Noise figure & $N_{f}$ & $5$ dB\\
   Frame length & $T_{F}$ & $10$\\
   Broadband user source block length & $O$ & $32$\\
   \Gls{iot} user packet length & $L$ & $128\,\textrm{B}$\\
   \Gls{iot} user activation probability & $p_{a}$ & $0.1$\\
   Path loss exponent & $\eta$ &$2.6$\\
	\bottomrule[0.4mm]	\end{tabular}\vspace{-0.4cm}
\end{table}
\section{Results}\label{NumRes}
In this section, we evaluate the performance of the proposed \gls{rl}-based policy optimisation approach for both the \gls{ran} Slicing and {RAN Sharing} 
scenarios. The broadband user is located at a distance $d_{b}\in \{35,75\}$\,m from the \gls{bs}, while the \gls{iot} users are placed within $d_{j}\in\{100,400\}$\,m. The results are averaged over $100$ independent simulations, where users are randomly positioned within their designated ranges in each run. Each simulation consists of at least $100000$ frames. 
For the broadband user, the transmission rate $r_{b}$ is selected as the minimum of the  maximum data rate corresponding to $P_{\textrm{max}}$, denoted by $r_{b}^{\textrm{max}}$, and the maximum achievable data rate satisfying the target error probability $\epsilon_{b}^{*}$, given by
\begin{equation}\label{eq:BBuser_r1}
    r_{b}=\max\{r\in (0,r_{b}^{\textrm{max}}]:\epsilon_{b,w}\left(r\right)=\epsilon_{b}^{*}, P_{b,t}\leq P_{\textrm{max}}\}.
\end{equation}
Following \eqref{eq:BBuser_r1}, $P_{b,t}$ can be calculated by assuming absence of interference in \eqref{eq:sinr_m} and then utilising error probability and threshold rate expressions, and is expressed as\cite{mishra2024coexist,Ley23Asilomar}
\begin{equation}\label{eq:Power_BBUser}
P_{b,t}=\min\left(\frac{\left(2^{r_{b}/B_{w}}-1\right)\sigma_{w}^{2}}{\mathbb{E}[|h_{b}|^{2}]\log(\epsilon_{b}^{*}-1)},P_{\textrm{max}}\right),
\end{equation}
assuming that the broadband user has statistical knowledge of its channel, i.e., $\mathbb{E}[|h_{b}|^{2}]$. We denote by $F(K)$ the random variable representing the number of frames needed to successfully decode a block of $K$ source packets, Subsequently, the throughput of the broadband user is calculated as
\begin{equation}\label{eq:BBuser_Throughput}
    S_{b}=r_{b}K/\left(\mathbb{E}\{F(K)\}T_{F}\right),
\end{equation}
and the \gls{ee} is calculated as $S_{b}/P_{b,t}$. Finally, the rest of the simulation parameters are given in Table \ref{tab:Sim_table}.   
\begin{figure}[!t]
    \centering
    \subfloat[\gls{ran} Slicing]{\includegraphics[width=0.5\linewidth]{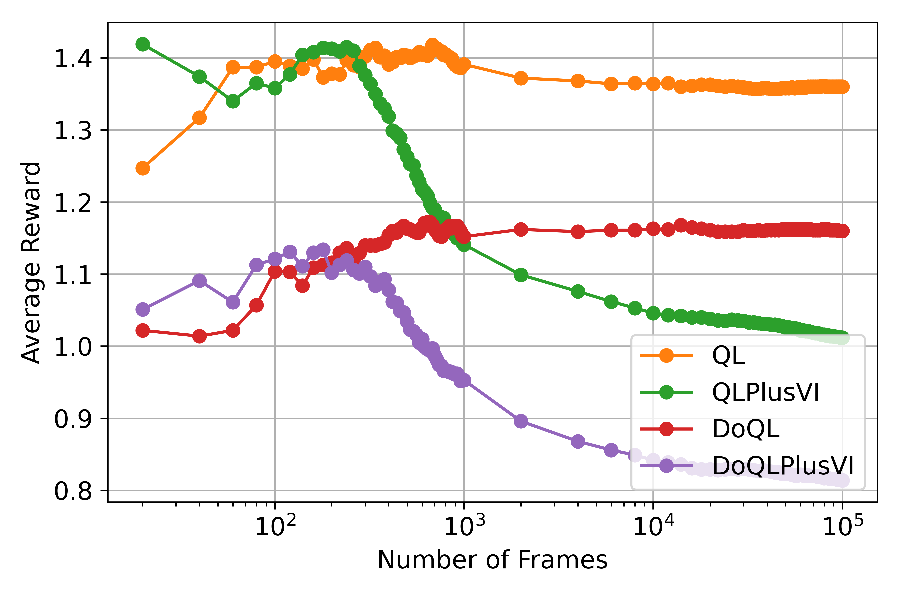}}
    \subfloat[\gls{ran} Sharing]{\includegraphics[width=0.5\linewidth]{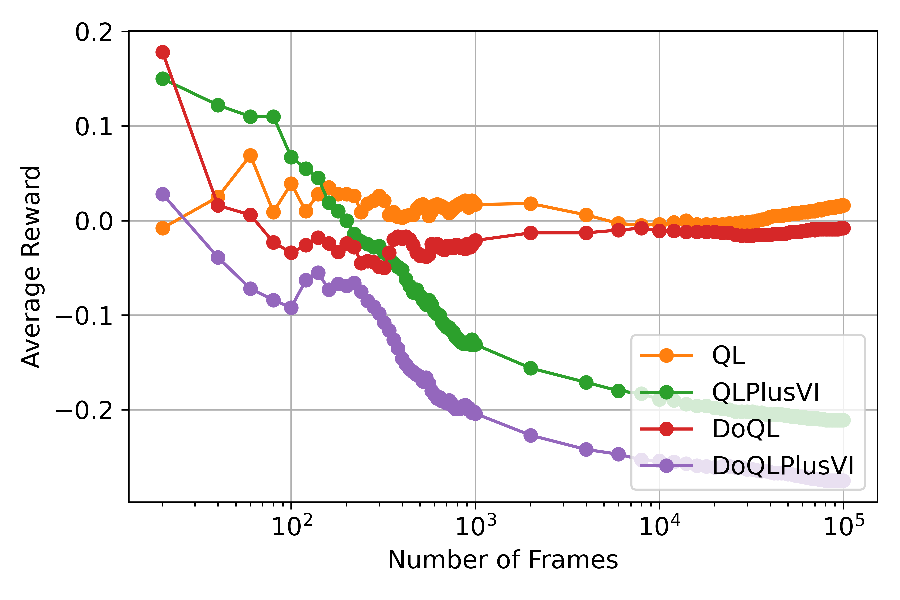}}
    \caption{Average rewards per packet during training phase, $J=10$}\vspace{-0.1cm}
    \label{fig:Training_with_Episodes}
\end{figure} 
\begin{figure}[!t]
    \centering
    \vspace{-0.6cm}\subfloat[$J=4$]{\includegraphics[width=0.5\linewidth]{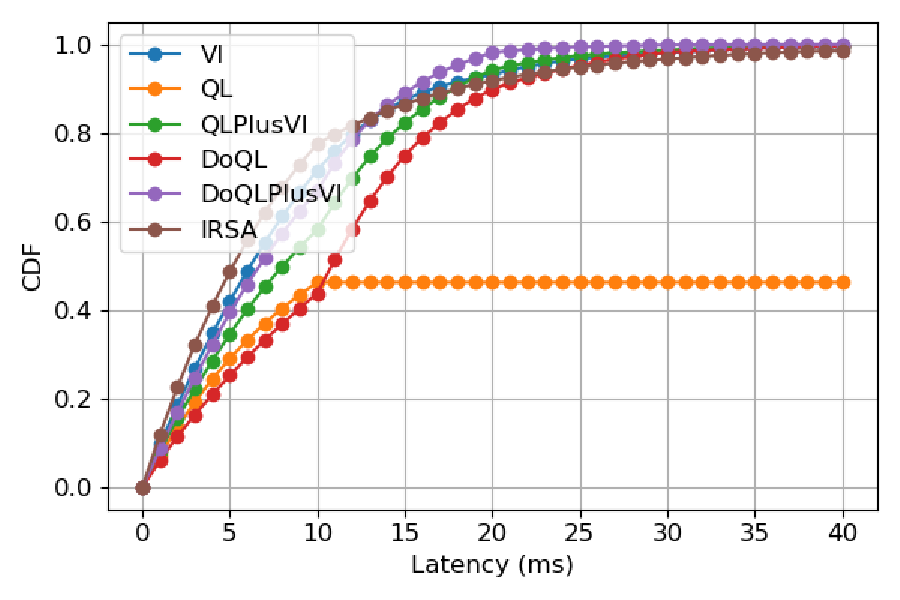}}
    \subfloat[$J=10$]{\includegraphics[width=0.5\linewidth]{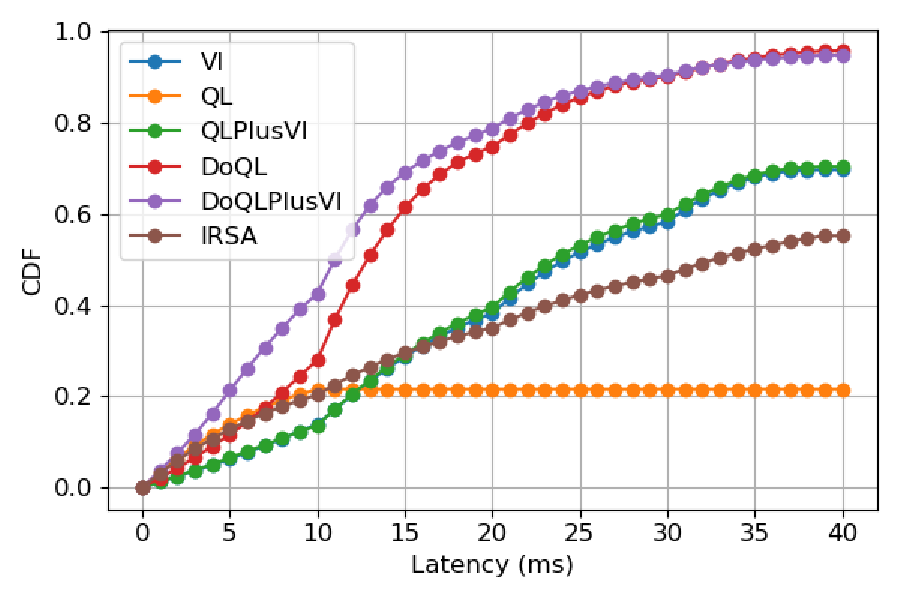}}
    \caption{Latency (ms) performance of \gls{iot} users with \gls{ran} Slicing}
    \label{fig:RANS_Latency}
\end{figure}
\begin{figure}[!t]
    \vspace{-0.5cm}\centering
    \subfloat[$J=4$]{\includegraphics[width=0.5\linewidth]{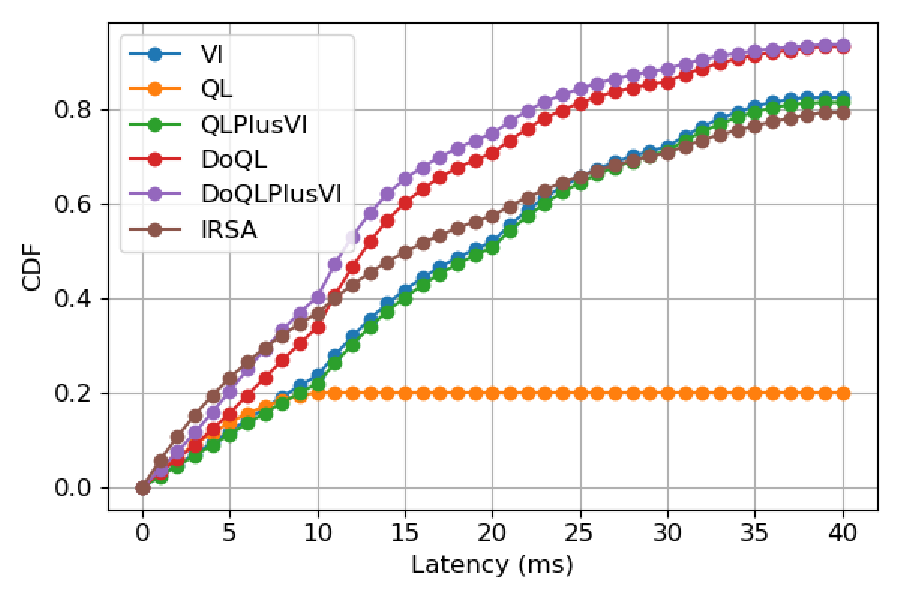}}
    \subfloat[$J=10$]{\includegraphics[width=0.5\linewidth]{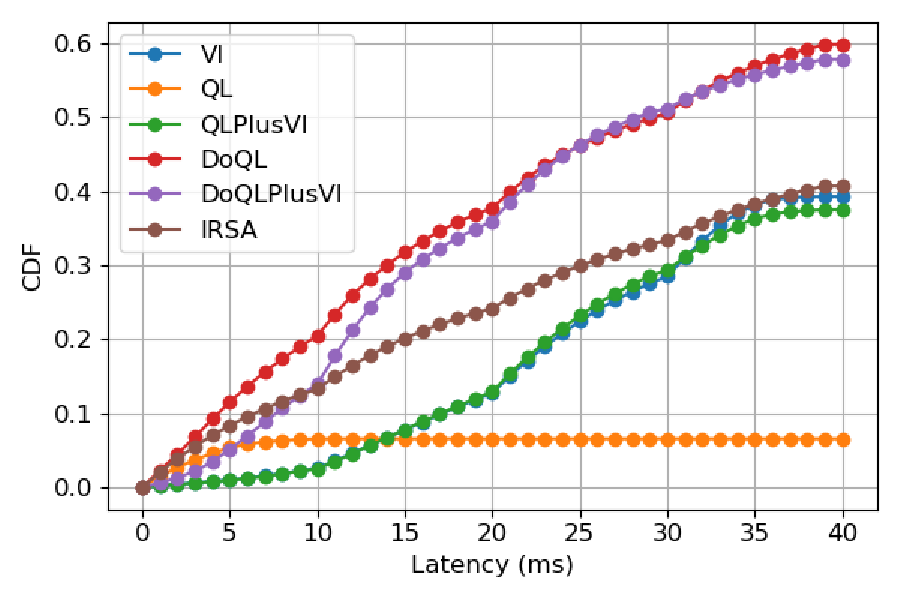}}
    \caption{Latency (ms) performance of \gls{iot} users with \gls{ran} Sharing}\vspace{-0.5cm}
    \label{fig:NoSL_Latency}
\end{figure}
\par Next, we illustrate and discuss the {latency performance} of the \gls{iot} users under the proposed optimisation framework. The schemes considered for performance evaluation and analysis are as follows: \textbf{1) VI} -- the policy of an \gls{iot} user is derived under the assumption that no other user is contending for access\cite{Ley23Asilomar}; \textbf{2) QL} -- the policy is obtained using the method described in Section~\ref{Perf_Opt}, but employing a single $Q$-function; \textbf{3) QLPlusVI} -- the policy is derived as in \gls{ql}, but the $Q$-function is initialised using the value function obtained from the \gls{vi} approach; \textbf{4) \gls{doql}} -- the policy is obtained using the \gls{doql} approach as described in Section~\ref{Perf_Opt}; \textbf{5) DoQLPlusVI} -- the policy is derived as in \gls{doql}, with the $Q$-function initialised based on the \gls{vi} approach, and \AM{\textbf{6) \gls{irsa}}-- the repetition degree distribution $\Lambda=0.25z^{2}+0.60z^{3}+0.15z^{8}$ proved to be superior to other commonly used
distributions\cite{Liva@DD,Eleni@DD}.} \AM{We begin with Fig.~\ref{fig:Training_with_Episodes}, which illustrates the evolution of the average rewards per packet of $J=10$ \gls{iot} users during the training phase.} The plot shows the rewards averaged over $5$ randomised user deployments. We avoid averaging over $100$ runs here, as it would overly smooth the curve. Nevertheless, the plot clearly demonstrates that convergence generally occurs within $5000$ frames. Accordingly, we adopt $5000$ frames as a reasonable training length across all approaches and scenarios, corresponding to a $5\%$ training cost. It can be observed that \gls{doql} is more robust to overestimation bias than \gls{ql} in stochastic environments, such as the case with $J=10$ users, under both \gls{ran} Slicing and \gls{ran} Sharing. Furthermore, in QLPlusVI and DoQLPlusVI, the rewards drop during training phase. This is due to the adjustment of the VI-initialised $Q$-functions to high-interference conditions. Note that the observed training phase rewards are not indicative of the inference performance; they are presented solely to analyse convergence period.
\begin{figure}
    \centering
    \subfloat[\gls{ran} Slicing]{\includegraphics[width=0.5\linewidth]{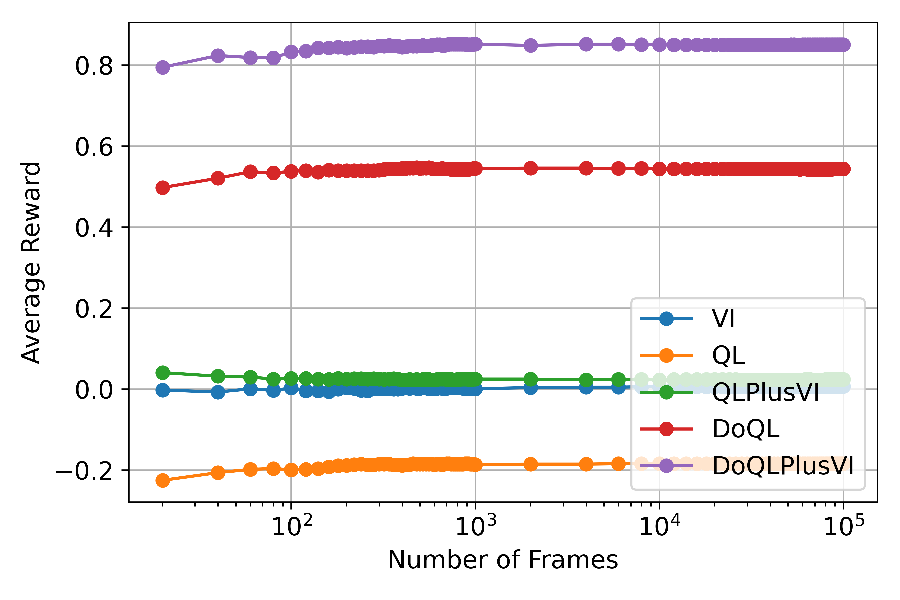}}
    \subfloat[\gls{ran} Sharing]{\includegraphics[width=0.5\linewidth]{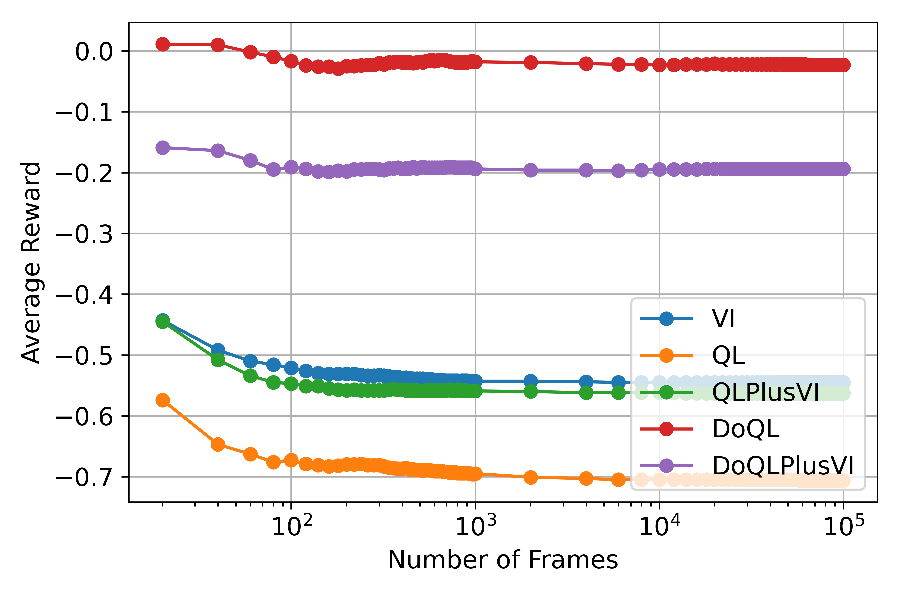}}
    \caption{Average rewards per packet during inference phase, $J=10$}
    \label{fig:NoSL_RANS_Reward}
\end{figure}
\begin{figure}
    \vspace{-0.5cm}\centering
    \subfloat[\gls{ran} Slicing]{\includegraphics[width=0.5\linewidth]{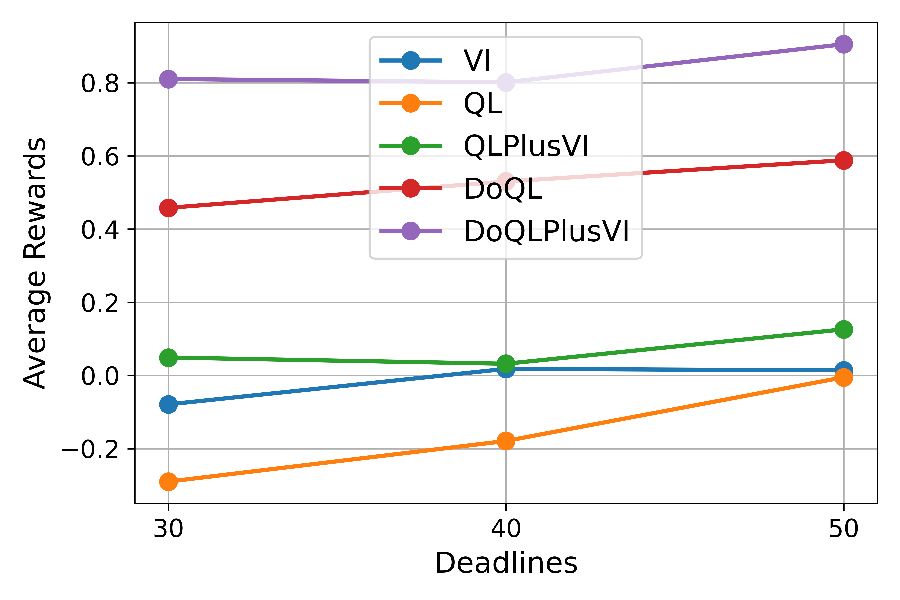}}
    \subfloat[\gls{ran} Sharing]{\includegraphics[width=0.5\linewidth]{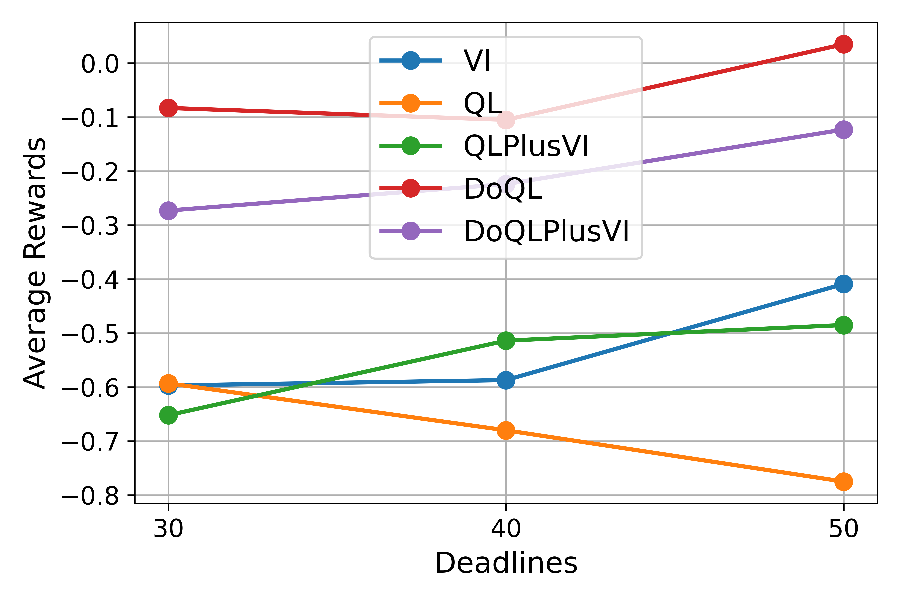}}
    \caption{Average rewards for different latency thresholds, $J=10$}\vspace{-0.6cm}
    \label{fig:RANS_NoSL_DL}
\end{figure}
\par Following this, Fig.~\ref{fig:RANS_Latency} and Fig.~\ref{fig:NoSL_Latency} illustrate the \gls{cdf} of the latency experienced by \gls{iot} users during inference phase under \gls{ran} Slicing ($B_{2}=B/2$) and \gls{ran} Sharing scenarios, respectively. In Fig.~\ref{fig:RANS_Latency}(a), for $J = 4$, the \gls{vi} approach outperforms \gls{doql} and performs comparably to DoQLPlusVI. This is due to the low and intermittent activation of a small number of users, which results in limited environmental variation, reducing the advantage of model-free approaches over the model-based \gls{vi}. \AM{The small $J$ also leads to the conventional \gls{irsa} policy achieving  a performance comparable to DoQLPlusVI.} As $J$ increases to $10$, both \gls{doql} and DoQLPlusVI outperform other baseline methods, with only these achieving the latency target with desirable reliability. In the \gls{ran} Sharing scenario, i.e. Fig.~\ref{fig:NoSL_Latency}, the continuous presence of the broadband user induces persistent interference, allowing \gls{doql} and DoQLPlusVI to outperform all others even at lower $J$. However, for $J = 10$, even these methods fail to meet the latency target with high reliability, albeit achieving significant performance gain over other schemes.\footnote{Note that while a reliability of $60\%$ may be insufficient for latency-critical applications, it can still be acceptable for goal-oriented objectives such as reconstruction error or actuation cost in source reconstruction scenarios~\cite{pappas@dtmcTWC}.} Furthermore, under such high-interference conditions, the benefit of initialising the $Q$-function using \gls{vi} diminishes and may even degrade performance, as the \gls{vi}-based policy implicitly assumes an interference-free environment. \AM{This is first evident in the shift in relative performance between DoQL and DoQLPlusVI as the number of \gls{iot} users increases from $J = 4$ to $J = 10$. The significant drop in the performance of both \gls{vi} and QLPlusVI further supports this observation, with performance  falling below than that of the \gls{irsa} policy with $J=10$. This degradation also explains the evolution of their rewards during the training phase, as illustrated in Fig.~\ref{fig:Training_with_Episodes}.} The rewards achieved by different \gls{rl} approaches during inference phase in Fig.~\ref{fig:NoSL_RANS_Reward} (for $J = 10$) align with the above observations. Furthermore, the relative performance among the \gls{rl} approaches remains consistent across different latency deadlines, as illustrated in Fig.~\ref{fig:RANS_NoSL_DL}.
\begin{figure}
    \vspace{-1cm}\centering
    \includegraphics[width=0.5\textwidth]{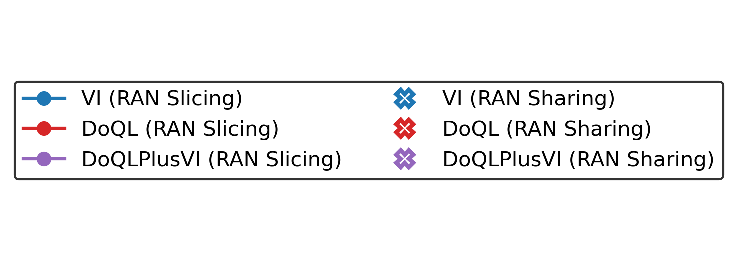}\vspace{-1.2cm}
    \hfill
    \subfloat[Throughput with $J=4$ \gls{iot} users]{\includegraphics[width=0.5\columnwidth]{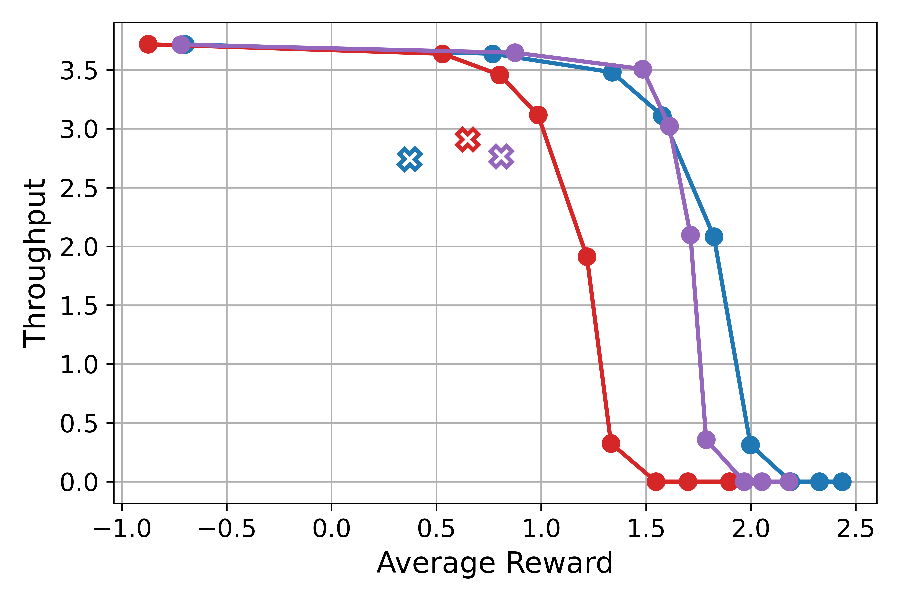}}
    \subfloat[\gls{ee} with $J=4$ \gls{iot} users]{\includegraphics[width=0.5\columnwidth]{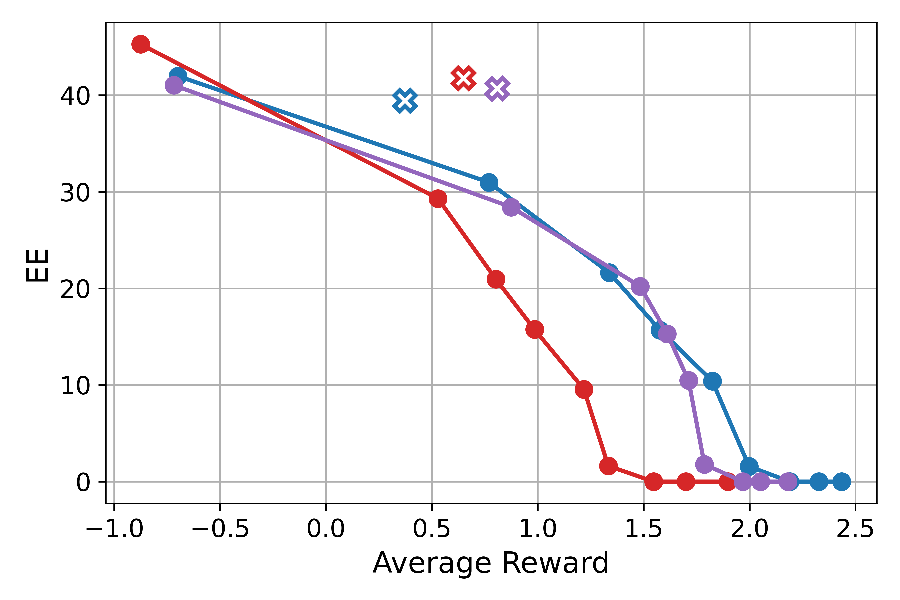}}
    \hfill
    \subfloat[Throughput with $J=8$ \gls{iot} users]{\includegraphics[width=0.5\columnwidth]{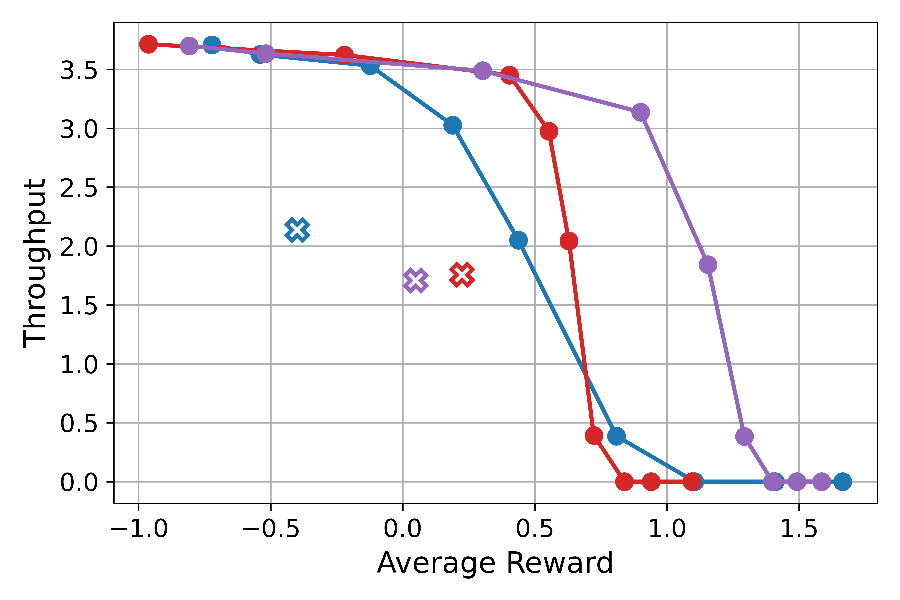}}
    \subfloat[\gls{ee} with $J=8$ \gls{iot} users]{\includegraphics[width=0.5\columnwidth]{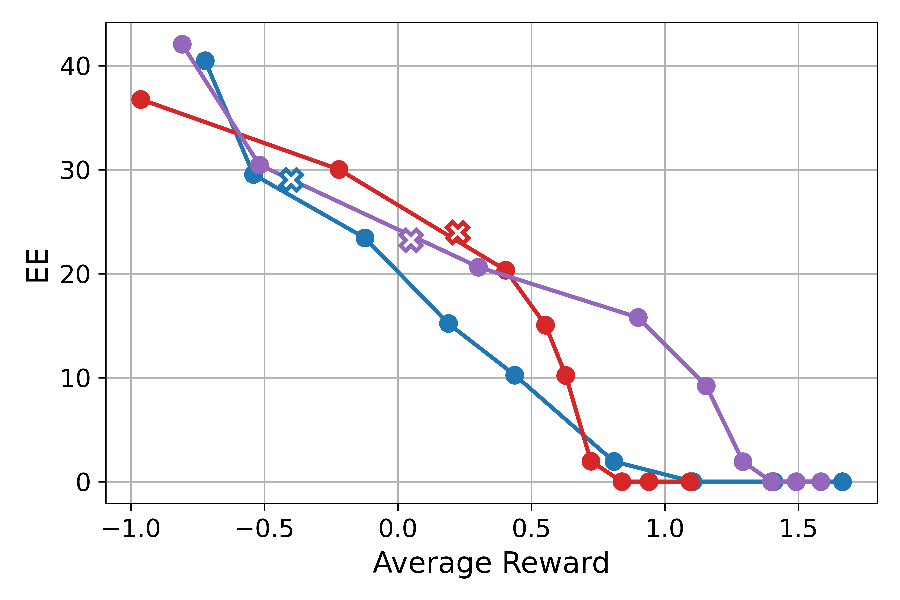}}
    \caption{Throughput and \gls{ee} of the broadband user versus average rewards of \gls{iot} users for \gls{ran} Slicing and \gls{ran} Sharing scenarios}\vspace{-0.6cm}
    \label{fig:Throughput_EE_Rewards}
\end{figure}
\par After evaluating and comparing various \gls{rl} approaches, we now examine the throughput and \gls{ee} performance of the broadband user when coexisting with \gls{iot} users under both \gls{ran} Slicing and \gls{ran} Sharing scenarios. Based on the earlier discussion, which established the superiority of \gls{doql} and DoQLPlusVI over other baseline schemes, respectively, we restrict our analysis to the \gls{vi}, DoQL, and DoQLPlusVI schemes. To this end, Fig.~\ref{fig:Throughput_EE_Rewards} presents the broadband user's throughput and \gls{ee} against the average rewards of \gls{iot} users for $J=4$ and $J=8$. In the \gls{ran} Slicing scenario, the operating points are obtained by varying the allocated bandwidth to \gls{iot} users, i.e., $B_{2} \in \{0.1B, \ldots, 0.9B\}$. As expected, lower values of $B_{2}$ lead to higher throughput and \gls{ee} for the broadband user, though this comes at the cost of significantly reduced rewards for \gls{iot} users. As $B_{2}$ increases, the average rewards achieved by the \gls{iot} users improve. Simultaneously, the throughput of the broadband user experiences only a marginal decline, as its transmit power $P_{b,t}$ can be adjusted to maintain the desired rate; see equations \eqref{eq:BBuser_r1} and \eqref{eq:Power_BBUser}. This trade-off is more evidently reflected in the steeper degradation of \gls{ee} performance, as illustrated in Fig.~\ref{fig:Throughput_EE_Rewards}(b) and Fig.~\ref{fig:Throughput_EE_Rewards}(d). For a small number of \gls{iot} users, i.e., $J=4$, Fig.~\ref{fig:Throughput_EE_Rewards} shows that both throughput and \gls{ee} versus average reward curves are nearly identical for \gls{vi} and DoQLPlusVI, while lower for DoQL. This suggests that, under such low-load conditions, learning offers limited benefit in the \gls{ran} slicing scenario. On the other hand, for a larger number of \gls{iot} users (i.e., $J=8$), the performance gains of \gls{doql} and DoQLPlusVI over \gls{vi} become clearly evident, even when higher bandwidths are allocated to the \gls{iot} slice. Finally, as $B_{2}$ approaches $B$, both the throughput and \gls{ee} of the broadband user degrade significantly. This is because, with a substantial increase in $B_{2}$, the remaining bandwidth $B_{1}$ available to the broadband user becomes severely limited. Consequently, even with $P_{\textrm{max}}$, the broadband user is unable to sustain the desired data rate $r_{b}^{\textrm{max}}$ while maintaining the required error probability.
\par For the \gls{ran} Sharing scenario, the throughput versus average reward performance yields a single operating point per \gls{rl} approach, as both services share the entire bandwidth, i.e., $B_3 = B$. As expected, all three \gls{rl} approaches exhibit degraded throughput performance under \gls{ran} Sharing when compared to their respective performances in the \gls{ran} Slicing scenario, for both $J = 4$ and $J = 8$. Nevertheless, \gls{doql} and DoQLPlusVI achieve higher average rewards than \gls{vi} while maintaining comparable throughput and \gls{ee}, highlighting the benefits of learning-based approaches in shared-bandwidth environments. Furthermore, for $J = 4$, the \gls{rl} approaches achieve higher \gls{ee} for the same average reward levels of the \gls{iot} users under \gls{ran} Sharing than under \gls{ran} Slicing. This suggests that in low-user scenarios, learning-based methods can enable a more favourable trade-off between the two services when operating under the \gls{ran} Sharing configuration. However, this advantage diminishes as the number of \gls{iot} users increases to $J = 8$, resulting in a high-interference environment. In such cases, \gls{ran} Slicing becomes essential to preserve the performance of both services.
\vspace{-0.15cm}
\section{Conclusion}\label{Concl}
This work investigated the coexistence of latency-constrained \gls{iot} users and a broadband service under both \gls{ran} Slicing and \gls{ran} Sharing configurations. We proposed a \gls{rl} framework based on \gls{doql} to optimise repetition-based access policies for the \gls{iot} users. Through numerical results, we demonstrated that the proposed approaches significantly outperform baseline methods in meeting latency target, in both \gls{ran} Slicing and \gls{ran} Sharing scenarios. Furthermore, in low \gls{iot} traffic conditions, the proposed schemes enable higher broadband user throughput with \gls{ran} Slicing and higher \gls{ee} with \gls{ran} Sharing. On the other hand, under high \gls{iot} traffic, the proposed schemes favour for higher throughput while maintaining \gls{ee} comparable to RAN Sharing. These findings highlight the potential of scalable decentralised \gls{rl} for efficient resource sharing and latency-aware scheduling in future multi-service wireless networks.
\vspace{-0.2cm}
\ifCLASSOPTIONcaptionsoff
  \newpage
\fi
\bibliographystyle{IEEEtran}
\bibliography{references}
\end{document}

%% file: acronyms.tex
\newacronym{mimo}{MIMO}{multiple-input multiple-output}
\newacronym{mmimo}{mMIMO}{massive MIMO}
\newacronym{ap}{AP}{access point}
\newacronym{qos}{QoS}{quality-of-service}
\newacronym{kpi}{KPI}{key performance indicator}
\newacronym{iot}{IoT}{internet-of-things}
\newacronym{aoi}{AoI}{age-of-information}
\newacronym{voi}{VoI}{Value-of-Information}
\newacronym{aoii}{AoII}{age-of-incorrect information}
\newacronym{ran}{RAN}{radio access network}
\newacronym{fec}{FEC}{forward error correction}
\newacronym{bbu}{BBU}{baseband unit}
\newacronym{ppp}{PPP}{Poisson point process}
\newacronym{cdf}{CDF}{cumulative distribution function}
\newacronym{dtmc}{DTMC}{discrete-time Markov chain}
\newacronym{paoi}{PAoI}{peak AoI}
\newacronym{nors}{NORS}{non-orthogonal RAN slicing}
\newacronym{ors}{ORS}{orthogonal RAN slicing}
\newacronym{fdma}{FDMA}{frequency-division multiple access}
\newacronym{noma}{NOMA}{non-orthogonal multiple access}
\newacronym{oma}{OMA}{orthogonal multiple access}
\newacronym{mse}{MSE}{mean squared error}
\newacronym{sic}{SIC}{successive interference cancellation}
\newacronym{bs}{BS}{base station}
\newacronym{isac}{ISAC}{integrated sensing and communications}
\newacronym{jsac}{JSAC}{joint sensing and communications}
\newacronym{rl}{RL}{reinforcement learning}
\newacronym{mdp}{MDP}{Markov decision process}
\newacronym{pomdp}{POMDP}{partially-observable MDP}
\newacronym{cpu}{CPU}{central processing unit}
\newacronym{tdd}{TDD}{time-division duplexing}
\newacronym{csi}{CSI}{channel state estimation}
\newacronym{urllc}{URLLC}{ultra-reliable low latency communications}
\newacronym{rsma}{RSMA}{rate-splitting multiple access}
\newacronym{sdma}{SDMA}{space division multiple access}
\newacronym{se}{SE}{spectral efficiency}
\newacronym{semcom}{SemCom}{semantic and goal-oriented communications}
\newacronym{6g}{6G}{sixth-generation}
\newacronym{5g}{5G}{fifth-generation}
\newacronym{ai}{AI}{artificial intelligence}
\newacronym{ml}{ML}{machine learning}
\newacronym{marl}{MARL}{multi-agent RL}
\newacronym{sarl}{SARL}{single-agent RL}
\newacronym{ack}{ACK}{acknowledgement}
\newacronym{nack}{NACK}{negative ACK}
\newacronym{cscg}{CSCG}{circularly symmetric complex Gaussian}
\newacronym{sinr}{SINR}{signal-to-interference plus noise ratio}
\newacronym{ee}{EE}{energy efficiency}
\newacronym{vi}{VI}{value-iteration}
\newacronym{ql}{QL}{Q-Learning}
\newacronym{doql}{DoQL}{Double QL}
\newacronym{v2x}{V$2$X}{vehicular-to-everything}
\newacronym{irsa}{IRSA}{irregular repetition slotted ALOHA}
\newacronym{plr}{PLR}{packet loss rate}